\documentclass[letterpaper, 10 pt, conference]{ieeeconf}  

\IEEEoverridecommandlockouts                              
\usepackage{graphicx} 
\usepackage{epsfig} 
\usepackage{amsmath} 
\usepackage{bbold}

\def\(({\left(}
\def\)){\right)}
\def\[[{\left[}
\def\]]{\right]}
\newcommand{\be}{\begin{equation}}
\newcommand{\ee}{\end{equation}}
\newcommand{\bea}{\begin{eqnarray}}
\newcommand{\eea}{\end{eqnarray}}

\usepackage{algorithm}
\usepackage[noend]{algpseudocode}
\usepackage{amsfonts}
\usepackage{flushend}

\title{\LARGE MMSE of probabilistic low-rank matrix estimation: \\Universality with respect to the output channel.}

\author{
Thibault Lesieur$^{1}$, Florent Krzakala$^{2}$, and Lenka Zdeborov\'a$^{1}$
\thanks{$^{2}$ LPS and CNRS, Ecole Normale Sup\'erieure, Rue Lhomond,
  Paris, France.}%
\thanks{$^{1}$ IPhT, CEA Saclay, and CNRS, 91191 Gif-sur-Yvette, France.}%
}

\begin{document}
\maketitle

\begin{abstract}
 This paper considers probabilistic estimation of a low-rank matrix from
 non-linear element-wise measurements of its elements.  We derive
 the corresponding approximate message passing (AMP) algorithm and its state
 evolution. Relying on non-rigorous but standard assumptions motivated
 by statistical physics, we characterize the minimum mean squared error (MMSE) achievable
 information theoretically and with the AMP algorithm. Unlike in
 related problems of linear estimation, in the present setting the MMSE
 depends on the output channel only trough a single parameter -- its Fisher information. We illustrate this striking finding by analysis
 of submatrix localization, and of detection of communities hidden in a
 dense stochastic block model. For this example we locate the
 computational and statistical boundaries that are not equal for rank
 larger than four. 
\end{abstract}

\section{Introduction}




Estimation of low-rank matrices from their noisy or incomplete
measurements is a problem that has a wide range of applications of practical
interest \cite{IgorBlog}. As for every broadly relevant data processing problem it is
of interest to study statistical and computational limits of such an
estimation on meaningful model settings. In this paper we evaluate the Bayes optimal and computationally
achievable mean-squared error of estimation for the following two models:

In the first model the matrix to be estimated is created
as 
\begin{equation}
    W=\frac{1}{\sqrt{n}} X K X^T \, ,\label{modelxx}
\end{equation}
where $X$ is a $n \times r$ matrix whose rows $x_i$ were chosen independently
at random from some distribution $P_{\rm prior}(x_i)$, and $K$ is a
$r\times r$ symmetric matrix. The matrix $W$ is then observed element-wise trough a 
noisy non-linear output channel $P_{\rm out}(y_{ij}|w_{ij})$, with $i,j=1,\dots,n$. The goal is to estimate the
unknown matrix $X$ from measured $Y$. 

We consider the problem in the limit of
very large systems $n\to \infty$, small rank $r=\Omega(1)$, and  $\partial
P_{\rm out}/\partial w=\Omega(1)$. The purpose of the scaling factor $1/\sqrt{n}$
in (\ref{modelxx}) is that the inference problem is neither trivially easy nor
clearly impossible in this limit. The same model and scaling was
considered e.g. in \cite{deshpande2014information,lesieur2015phase}, where the matrix $K$ was an identity.  
The main algorithmic difficulty in such a setting comes from the
high dimension $n$. In this work we study the idealized setting in which the hyper-parameters $r, K, P_{\rm prior},
P_{\rm out}$ are independent of the dimension $n$ and known. Note,
however, that extending our approach via expectation
maximization seems like a natural way to learn these hyper-parameters.

In the second model the matrix to be estimated is created as 
\begin{equation}
    W=\frac{1}{\sqrt{n}} U V^T \, , \label{modeluv}
\end{equation}
where $U$ is a $n \times r$ matrix and $V$ is a $m \times r$ matrix,
with rows $u_i$ (resp. $v_i$) chosen from some distribution
$P^u_{\rm prior}(u_i)$ (resp. $P^v_{\rm prior}(v_i)$). The matrix $W$ is then
observed trough the same output channel $P_{\rm out}(y_{ij}|w_{ij})$ as above,
and the problem is set analogously, adding that
$m/n=\alpha=\Omega(1)$. 
Most of the discussion in this paper uses notation of the first model,
but all our results are relevant for the second one as well. 

An algorithmic tool that first comes to our mind when seeing the setting
above is singular (or eigenvalue) value decomposition (SVD) keeping the leading $r$
components. SVD is the optimal estimation algorithm if the goal is to minimize
the mean squared distance between the observed matrix $Y$ and the estimator irrespectively of the properties of the factors (such as the prior
$P_{\rm prior}$). Requirements on the factors, such as sparsity or
other structure are not incorporated in the basic SVD. Moreover, for a
general non-linear output channel $P_{\rm out}$  the sum
of squares between $Y$ and its estimator is not the most relevant quantity to
minimize.

Setting the problem in a fully probabilistic way is another common
approach \cite{mnih2007probabilistic},
that is in principle much more flexible, but algorithmically more
challenging in general. To obtain a Bayes-optimal estimator of the factor
$x_i$ ($r$-dimensional column vector) we need to compute the marginals of the posterior probability
distribution
\begin{equation}
   P(X|Y) =   \frac{1}{Z(Y)} \prod_{1 \le i \le n}  P_{\rm prior}(x_i)  
      \prod_{1 \le i \le j \le n} P_{\rm out}(y_{ij}|w_{ij})
      \, , \label{eq:post}
\end{equation}
where $w_{ij}=x^T_i K x_j/\sqrt{n}$. In this paper we will leverage this
algorithmic difficulty by realizing that techniques based on
approximate message passing and related state evolution (SE) are
asymptotically optimal for the above setting. For compressed sensing
these techniques are rigorous thanks to series of works
\cite{DonohoMaleki09,BayatiMontanari10}. The AMP algorithm and its
state evolution for the above setting is different from the one
of compressed sensing, and these proofs do not apply. Arguments for the
optimality in the present setting come from non-rigorous methods of statistical
physics. From a mathematical point of view the present paper provides a
set of accurate conjectures. Fully rigorous proof of these
conjectures is a natural direction for future work.

\subsection{Examples}

The above model includes a number of examples that are commonly
considered in the literature. Without trying to be exhaustive, we list
a few interesting ones. 

\begin{itemize}
  \item{Sparse principle component analysis as considered in
      \cite{deshpande2014information,lesieur2015phase}: The prior distribution
      $P_{\rm prior}$ has a weight $1-\rho$ on a vector of zeros. The
      output channel was considered as additive Gaussian noise. }
  \item{Robust PCA \cite{wright2009robust} in which the measured matrix is a low-rank
      matrix plus a sparse large noise: The output channel adds small noise
      with some probability $p$ and large noise with $1-p$.}
  \item{Submatrix localization \cite{shabalin2009finding,ma2015computational}: The matrix $W$ includes $r$
      submatrices (overlapping or not) that have larger mean than the
      overall mean of the matrix $W$. The prior $x_{i}$ then encodes
      in a binary manner to which of the $r$ submatrices does a given
      variable $i$ belong. The output is usually considered as Gaussian additive noise.}
  \item{Detection of communities hidden in dense networks: Stochastic block model is
      popular for theoretical studies of clustering. Nodes belong to
      different clusters/communities, the prior $P_{\rm prior}$ then
      allows only vectors having one component $1$, and $0$ elsewhere. The
      observed matrix $Y$ is binary and the probability to observe
      $y_{ij}=1$ is given by $K_{ab}$ if $x_i(a)=1$ and $x_j(b)=1$
      ($i,j=1,\dots,n$, and $a,b=1,\dots,r$).}
  \item{Biclustering is a simultaneous clustering of rows and columns,
      it finds a number of applications e.g. for analysis of microarray data in
    genomics \cite{cheng2000biclustering}. Again the prior encodes
    affinity to a cluster and the output function includes various models
    of noise.}
  \item{Poisson noise in the matrix factorization was considered
      e.g. in \cite{gopalan2013scalable}.}
 \item{The labeled stochastic block model is another case that can be
     reformulated in the present setting \cite{lelarge2013reconstruction}.}
\end{itemize}

\subsection{Contribution and closely related work}

As far as we know the only tools that provides asymptotically exact
analysis of the minimal mean squared error of
models (\ref{modelxx}-\ref{modeluv}) is approximate message passing and
state evolution as deployed in the present paper. For the output channel being additive
Gaussian noise this was done previously in
\cite{rangan2012iterative,deshpande2014information} for rank $r=1$
with part of the results being fully rigorous, in \cite{matsushita2013low}
for generic rank and without the state evolution, and in \cite{lesieur2015phase} for general rank with the state
evolution, but non-rigorously. The main contribution of this paper is
the treatment of the case of general non-linear output channel $P_{\rm out}$. 

Approximate message passing and state evolution for a generic output
channel $P_{\rm out}$ was derived previously in the context of linear estimation
\cite{Rangan10b}, and later in matrix factorization with $r=\Omega(n)$ \cite{kabashima2014phase}. In
both these cases the resulting equations (both the AMP and the state evolution) are considerably more
involved than those for additive Gaussian noise. In the setting of
low-rank models (\ref{modelxx}-\ref{modeluv}) above the situation is
remarkably simpler as the AMP algorithm stays the same up to a change
of the matrix $Y$ for the so-called Fisher score matrix $S$ that depends on the
output channel and on $Y$ element-wise, and the inverse of the Fisher
information of the channel that we denote $\Delta$ and plays a role
of an effective noise variance. In the state evolution
the situation is even simpler in the sense that only the effective
value $\Delta$ of the noise appears. 

The space of all the possible
element-wise output channels hence reduces to one dimensional space,
parametrized by their inverse Fisher information $\Delta$. The resulting asymptotic MMSE
depends only on $\Delta$ and not on other details of the
channel. Consequently, classes of rather differently looking matrix
estimation problems all share the same single-letter characterization.  The contribution of the
present paper is to unveil and quantify this property, and illustrate it on the
example of detection of communities hidden in dense networks that is
via this one parameter mapping related to localization of submatrices
having different mean from the background matrix.

Analogous universality with respect to the output channel was observed
in \cite{deshpande2015finding} (see e.g. their remark 2.5) in the
study of detection of a small hidden clique with approximate message passing. 

Our results for the Bayes-optimal estimation error of community detection in dense stochastic block
model are of independent interest.  Analogous results were derived for
the sparse case in \cite{decelle2011asymptotic}. In the dense case
only MSE-suboptimal spectral methods were evaluated
\cite{nadakuditi2012graph}. We also unveil a hard phase existing
in this problem for rank $r>4$, and becoming very wide for $r\to
\infty$.  

We also recently learned about independently ongoing work of \cite{deshpande2015asymptotic} who consider
rank $r=1$ with Radamacher prior, and establish rigorously the relation between the
stochastic block model with two groups and low-rank estimation with
Gaussian channel.  

\section{AMP and state evolution}

\subsection{From belief propagation to AMP}
In this section we derive the AMP algorithm to compute marginals of
the posterior probability distribution (\ref{eq:post}). We present the
derivation for the $X K X^\top$ case, for the $UV^\top$ everything
works analogously and we only state the results.
For convenience we introduce a function $g(y,w)$ as 
\begin{equation}
P_{\rm out}(y|w)= e^{g(y,w)}\, .  \label{gout}
\end{equation}
We require  $g(y,w)$ to be differentiable in $w$. For the previously
considered  Gaussian additive noise we have $g_{\rm GAN}(y,w) = -(y-w)^2/(2 \Delta)$.

In the first step, we write the belief propagation equations for the
probability distribution (\ref{eq:post}). For this we introduce
messages ${m^t_{i j \rightarrow i}}(x_{i})$ and
$n_{i \rightarrow ij}^t(x_{i})$ between variables $x_i$ and the
factors associated to $y_{ij}$. The BP equations read
\begin{multline}
{m^t_{i l \rightarrow i}} (x_{i}) = \frac{1}{Z^t_{il \to i}} 
 \int {\rm d}x_l\,  {n^t_{l \rightarrow i l}}(x_l)\,  e^{g\left(y_{i
       l},\frac{1}{\sqrt{n}} x_{l}^{\top}K x_{i} \right)  } \,
 ,\\
{n^{t+1}_{i \rightarrow i j}} (x_{i}) = \frac{1}{Z^t_{i \to ij}} P_{\rm prior}(x_{i})
\prod\limits_{l \neq i,l \neq j} {m^t_{i l
    \rightarrow i}}(x_{i})\, , \label{Message2}
\end{multline}
where $Z^t_{il \to i}$ and $Z^t_{i \to ij}$ are normalizations.
The main assumption behind these BP equations is that the messages
$n_{i\rightarrow i l}(x_i)$  and  $n_{i\rightarrow i j}(x_i)$ can be
interpreted as probabilities that are
statistically independent conditioned to the values of the variable
$x_i$ in the large $n$ limit.  
This is also the main assumption of this paper. Arguments of
theoretical statistical physics \cite{MezardMontanari09} provide
heuristic justification for this assumption. 
In the above form the BP equations are not useful for implementation. We,
however, realize that we can rewrite them into a much simpler form that provides the same marginals in the limit
of large $n$. 

In the large $n$ limit the function $g\left(y_{i l},x_{l}^{\top}K x_{i}
  /\sqrt{n} \right)$ depends only weakly on $x_i$ and $x_l$. Therefore
we expand this function around $w=0$ and (\ref{Message2}) then becomes
\begin{multline}
\frac{{m^t_{i l \rightarrow i}} (x_{i})}{ \exp(g(y_{il},0))} \propto 1 + 	\frac{1}{\sqrt{n}}
 \frac{\partial g}{\partial w}\Big|_{y_{i l},0} {a_{l \rightarrow i l}^{t}}^\top    K x_{i} +
\\
\frac{1}{2n}   h(y_{il})  x_{i}^{\top}K\left(v_{l \rightarrow i l}^t +
a_{l \rightarrow i l}^t {a_{l \rightarrow i l}^{t}}^\top   
\right)
K x_{i} +  O\left({n^{-\frac{3}{2}}} \right) \, , \label{ExpansionBP}
\end{multline}
where 
\begin{equation}
  h(y_{il})=  \frac{\partial^2 g}{\partial w^2}\Big|_{y_{i l},0}
  + \left( \frac{\partial g}{\partial w}\Big|_{y_{i l},0} \right)^2\, .
\end{equation}
Here we introduced quantities $a^t_{l \rightarrow i l}$ and $v^t_{l \rightarrow i l}$ as the mean and covariance of  $n^{t}_{l \rightarrow i l}$
We then take the logarithm of (\ref{ExpansionBP}) and by using (\ref{Message2}) we find
\begin{equation}
{n ^{t+1}_{i \rightarrow i j}} (x_{i}) = P_{\rm prior}(x_{i})
e^{{B^{t}_{i \rightarrow
      ij}}^{\top}x_i - \frac{x_i^\top    A_{i \rightarrow ij}^t
    x_i}{2} }\, ,\label{Message4}
\end{equation}  
where 
\begin{align}
B^t_{i \rightarrow ij} = \frac{K}{\sqrt{n}}\sum\limits_{l \neq i,j} 	
\frac{\partial g}{\partial w}\Big|_{y_{i l},0} {a_{l \rightarrow i
    l}^{t}} \, , \\
A^t_{i \rightarrow ij} = \frac{K}{n}\Big\{ \sum\limits_{l \neq i,j} 	
\left( \frac{\partial g}{\partial w}\Big|_{y_{i l},0}\right)^2 a_{l
  \rightarrow i l}^t {{a_{l \rightarrow i l}^{t}}^\top} + \nonumber \\ 
 \sum\limits_{l \neq i,j}  h(y_{il})  \left(v_{l \rightarrow i l}^t +
a_{l \rightarrow i l}^t {a^t_{l \rightarrow i l}}^\top   
\right)  \Big\} K \,.\label{Define_A}
\end{align}
To close these equations we finally compute the new mean and variance of the messages $n_{i
  \rightarrow i j}^{t+1}$ as
\begin{eqnarray}
a_{i \rightarrow ij}^{t+1} = f(A_{i \rightarrow ij}^t,B_{i \rightarrow
  ij}^t)\, ,
\\
v_{i \rightarrow ij}^{t+1} = \left( \frac{\partial f}{\partial
    B}\right) (A_{i \rightarrow ij}^t,B_{i \rightarrow ij}^t)\, ,
\end{eqnarray}
where $f(A,B)$ is the mean of the normalized probability distribution
\begin{equation}
P(x)=\frac{1}{{\cal Z}(A,B)}P_{\rm prior}(x)\exp\left(B^\top    x - \frac{x^\top    A x}{2} \right)
\label{DefineFuncF}
\end{equation}
and $\partial f / \partial B$ is its covariance matrix. Above we closed the intractable distributional belief propagation
equations on the means and variances of the messages. 

As we can already anticipate it will be instrumental to introduce the
so-called Fisher score matrix (evaluated at $w=0$) as 
\begin{equation}
S_{ij} \equiv \frac{\partial \log P_{\rm out}(y_{ij}|w)}{\partial w}\Big|_{y_{ij},0}  \, ,\label{Y_eff}
\end{equation}
and the Fisher information (evaluated at $w=0$) of the output channel as 
\begin{equation}
   \frac{1}{\Delta} \equiv \mathbb{E}_{P_{\rm out}(y|w=0)}\left[
  \left( \frac{\partial \log P_{\rm out}(y|w)}{\partial w}\Big|_{y,0}\right)^2\right]\, . \label{Noise_Prop_3}
\end{equation}
We denote the inverse Fisher information by $\Delta$ because that would be the noise variance for
the Gaussian additive channel (in that case $S = Y/\Delta$). 

Using $\int {\rm d}y P_{\rm out}(y|w)=1$, $\forall w$ and (\ref{gout}), it follows that
\begin{align}
\mathbb{E}_{P_{\rm out}(y|w=0)}\left( \frac{\partial g}{\partial
    w}\Big|_{y,0} \right) = 0 \, ,\label{Noise_Prop_1}
\\
\mathbb{E}_{P_{\rm out}(y|w=0)}\left[ \left( \frac{\partial
      g}{\partial w}\Big|_{y,0}\right)^2 +  \frac{\partial^2
    g}{\partial w^2}\Big|_{y,0}\right] = 0\, . \label{Noise_Prop_2}
\end{align}
With the above, eqs. (\ref{Define_A}) can be rewritten as
\begin{align}
B^t_{i \rightarrow ij} = \frac{K}{\sqrt{n}}\sum\limits_{l \neq i,j} 	
S_{il} {a_{l \rightarrow i
    l}^{t}}  \, ,\\
A^t_{i \rightarrow ij} = \frac{K}{n\Delta} \left[ \sum\limits_{l \neq i,j} 	
a_{l
  \rightarrow i l}^t {{a_{l \rightarrow i l}^{t}}^\top} \right] K \, .\label{Define_As}
\end{align}
Here we can recognize AMP equations derived and studied for the low-rank matrix
estimation problem with additive Gaussian noise in
\cite{rangan2012iterative,matsushita2013low,deshpande2014information,lesieur2015phase}. 

Remarkably the non-linear output channel enters only trough the value of
the effective noise (\ref{Noise_Prop_3}) and an effective form of the observed
matrix (\ref{Y_eff}). This is much simpler than what happens for the
linear estimation model for which the generalized output AMP was
derived in \cite{Rangan10b}, or the matrix
factorization \cite{kabashima2014phase}. 

Finally we provide a simplification that is standard to AMP-like algorithms and
that could be called TAPyfication~\cite{ThoulessAnderson77}.
We notice that variables $a_{i \rightarrow ij}^t$ and $v_{i
  \rightarrow ij}^t$ depend only weakly on the index $j$, and this
allow us to reduce further the number of variables to end up with 
\begin{align}
& A^t = \frac{K}{n\Delta}\left[\sum\limits_{1 \leq l \leq n}
 a_l^t  {a_l^t}^\top\right]K \, , \label{Def_A_TAP}
\\
& B_i^t = \frac{K}{\sqrt{n}}\sum\limits_{l \neq i}
S_{il} {a_{l}^{t}} -
\frac{K}{\Delta} \left( \frac{1}{n}\sum\limits_{1 \leq l \leq n} v_l^{t} \right) K a_i^{t-1} \, ,\label{Def_B_TAP}
\\
& a_i^{t+1}= f(A^t,B_i^t)\, ,\label{new_ai}
\\
& v_i^{t+1}= \left(\frac{\partial f}{\partial B}\right)(A^t,B_i^t)\, .\label{new_vi}
\end{align}
Where the second term in the expression for $B_i^t$ is the so-called
Onsager reaction term, with its time index one iteration earlier, as
is usual in AMP-type algorithms. Here we have reduced the number of messages to iterate from $O(n^2)$ to $O(n)$.

The procedure carried out above to derive the AMP algorithms can also
be used to obtain a corresponding expression for the log-likelihood
$\phi= \log(Z(Y))/n$ where $Z(Y)$ is the normalization in
(\ref{eq:post}). This is called the Bethe free energy. 
Given a fixed point the AMP equations the Bethe free energy reads
(for details see Appendix B)
\begin{equation}
\phi = \frac{1}{n}\sum\limits_{1 \leq i \leq n} \log( {\cal Z}(A,B_i))-
\frac{1}{2n}\sum\limits_{1 \leq i \leq n} \log(\stackrel{\sim}{Z_i})
\, ,
\end{equation}
where ${\cal Z}(A,B_i)$ is the normalisation from (\ref{DefineFuncF}) and
{\small
\begin{equation}
\log(\stackrel{\sim}{Z_i})={\rm Tr}\left[\frac{a_i K}{\sqrt{n}}
  \sum\limits_{1 \leq j \leq n} S_{ij}a_j^\top    - \frac{K
    a_i a_i^\top  K }{\Delta n}\sum\limits_{1 \leq j \leq
    n}\frac{a_j a_j^\top}{2} + 2 v_j   \right] \, .
\end{equation}
}
This was derived in Appendix B.

\subsection{Summary of the algorithm}
For a given output channel we first need to evaluate the effective
noise parameter $\Delta$ (\ref{Noise_Prop_3}) and the score
matrix $S$ (\ref{Y_eff}). 
Then at each iteration of the AMP algorithm we store the following variables.
\begin{itemize}
\item $a_i^t$ and $a_i^{t-1}$ are the estimators of the mean of
the variables $x_i$ at time $t$ and $t-1$. Every $a_i$ is a vector of size $r \times 1$.
\item $v_i^t$ is the estimator of the covariance of
the variables $x_i$ at time $t$. The $v_i$ are matrices of size $r \times r$.
\end{itemize}
We then compute the matrix $A^t$ and the vectors $B_i^t$ with (\ref{Def_A_TAP}) and (\ref{Def_B_TAP}).
Every $B_i^t$ is a vector of size $r \times 1$ while
$A^t$ is a $r \times r$ matrix.
Finally we compute the new estimate of the mean and covariance of
variables $x_i$ from (\ref{new_ai}) and (\ref{new_vi}). 
Where the function $f$ is defined as mean and covariance of the
distribution in eq. (\ref{DefineFuncF}).
We initialize the $a_i^{t=0}$ very close to 0 in order
to avoid convergence problems.
To help convergence we can also damp the iterations with some
parameter~$\gamma$. More sophisticated damping should be implemented
if convergence problems persist on non-synthetic data \cite{vila2014adaptive}.
This finally gives us the following algorithm.
\begin{algorithm}[h!]
\caption{AMP for low-rank estimation}\label{euclid}
\begin{algorithmic}[1]
\State $X \gets 10^{-10}{\rm Normal}(n,r)$
\State $X_{\rm old} \gets {\rm Zeros}(n,r)$
\State $v \gets {\rm Zeros}(r,r)$
\State ${\rm diff} \gets 1$
\While {$t < t_{\rm max}$}
\State $B \gets \frac{1}{\sqrt{n}}S XK - \frac{X_{\rm old}Kv K}{n\Delta}$
\State $A \gets \frac{K X^{\top}XK }{\Delta n}$
\State $X_{\rm New} \gets f(A,B)$
\State $v_{\rm New} \gets\sum\limits_i \frac{\partial f}{\partial B_i}(A,B)$
\State $X_{\rm old} \gets X$
\State $X \gets (1-\gamma)X_{\rm New} +\gamma X$
\State $v \gets (1-\gamma)v_{\rm New} +\gamma v$
\State ${\rm diff} \gets {\rm norm}(X-X_{\rm old})^2/n$
\If{$t > t_{\rm min} \textbf{ and } {\rm diff} < 10^{-6}$}
\State $\textbf{break}$ 
\EndIf
\EndWhile
\State $\textbf{return } X$ 
\end{algorithmic}
\end{algorithm}

\subsection{Remark about spectral algorithms} 


An interesting remark can be done when noticing that there is an
intimate relation between AMP and SVD. First notice that for some prior distributions AMP has a
so-called uniform (or factorized) fixed point where the values of
$a_i$ do not depend on the index $i$ (e.g. for zero mean priors). 
In such cases it is meaningful to linearize AMP around this uniform
fixed point. This linearization can then be interpreted as a spectral
algorithm.  For the most common additive Gaussian output channel this
gives the eigen-decomposition of $Y$ (or SVD of $Y$ for the $uv^\top$ case). Of
course this spectral algorithm comes with its advantages
(non-parameteric) and disadvantages (hard to ensure constraints on the
factors $X$).

For a generic output channel the object that comes out from the AMP
linearization is the Fisher score matrix $S$ defined by
(\ref{Y_eff}). Following the analogy, the SVD decomposition (or eigen-decomposition) should hence be done on
$S$ rather than on $Y$. 
Let us give an example in which the difference between doing SVD on $Y$
or $S$ indeed improves performance of the spectral
method. We consider 
rank $r= 1$ and $P_{\rm prior}(x) = e^{-x^2/2}/ \sqrt{2 \pi}$.
The output channel is an additive exponential noise of
  parameter $\lambda$, i.e. 
$ P_{\rm out}(y|w) =\exp\left(-|y -   w|/\lambda\right)/(2 \lambda)$.
The score matrix $S$ is in this case proportional to 
$S \approx {\rm sign}(y_{ij})$. And indeed for
$1/\sqrt{2} < \lambda < 1$ an informative eigenvector gets out
of the bulk of the spectrum of $S$ but not out of $Y$'s
spectrum. Taking the score matrix allowed us to
extract meaningful eigenvectors.

\subsection{State evolution}
\label{SE}



Remarkably, the AMP algorithm is amenable to asymptotic analysis. In
statistical physics this would be called the cavity method
\cite{MezardMontanari09}, in linear estimation the commonly used term
is state evolution. The SE was derived for low-rank matrix
estimation for the Gaussian additive channel in \cite{rangan2012iterative,deshpande2014information,lesieur2015phase} and the present
case is a straightforward generalization.  To write the SE for the present case
we introduce two order parameters of dimension ${r \times r}$.
\begin{equation}
Q^t = \frac{1}{n}\sum\limits_{1 \leq i \leq n} a_i^t {a_i^t}^\top \, ,\label{op_Q}  \quad
M^t = \frac{1}{n}\sum\limits_{1 \leq i \leq n} a_i^t {x^t}_i^\top   \, ,
\end{equation}
where $X$ is the vector we are trying to infer and that is unknown to
us.
State evolution (or single letter characterization)
relies on the computation of the distribution (with respect to the
realizations of the matrix $Y$) of $B_i^t$ given $X$ ($A^t$ is a deterministic variable of
$Q^t$).

Using (\ref{Define_A}) we notice that $B_i^t$ is a sum of many terms
that by the assumptions of the belief propagation are
independent. Using the central limit theorem we can characterize
$B_i^t$ by its mean and variance as done in Appendix A. As a result one finds that
\begin{equation}
B_i^t = \frac{K M^t K {x}_i}{\Delta} + \xi_i \, ,
\end{equation}
where $\xi_i$ is a Gaussian random variable of mean 0 and of covariance
$K Q^t K/\Delta$. This allows us to see that the order
parameters $Q^t$ and $M^t$ will evolve as
\bea
&Q^{t+1} = \mathbb{E}_{P_x,P_\xi}
\left[ {f{(\frac{KQ^tK}{\Delta},\frac{K M^t K }{\Delta}x + \xi)}
    f^{\top   }{(.,.)}} \right],
\label{MSE_Q}
\\
&M^{t+1} =\mathbb{E}_{P_x,P_\xi}  \left[
{f(\frac{KQ^tK}{\Delta},\frac{K M^t K}{\Delta}  x+ \xi)} x^{\top   }\right],
\label{MSE_M}
\eea  
where we used abbreviations for $P_x =P_{\rm prior}(x)$ and $P_\xi={\cal
N}(0,KQ^tK/\Delta)$. The remarkable part of these equations
is that all the details of the output channel have disappeared and the
only thing that remains is the inverse of the Fisher information
$\Delta$. Larger $\Delta$ means larger effective noise, smaller Fisher
information and hence harder inference. The estimation error in the large $n$ limit depends on the channel only
trough its Fisher information.

Since in this paper we assume the knowledge of the generative model
and its parameters the state evolution simplifies further as we observe
that $Q^t = M^t$. This is called the Nishimori condition
in statistical physics and was derived e.g. in \cite{kabashima2014phase} or
\cite{deshpande2014information}.

State evolution also allows us to derive the Bethe free energy in
the large $n$ limit as 
\bea
     \phi &= \mathbb{E}_{P_x,P_\xi}  \left[ \log {\cal
         Z}(\frac{KQK}{\Delta},\frac{KM Kx}{\Delta}+\xi) \right]
\nonumber \\
&- \frac{1}{2 \Delta} {\rm Tr}(KMK M^{\top   }) + \frac{1}{4 \Delta} {\rm
  Tr}(KQK Q^{\top   }) \, . \label{Bethe_SE}
\eea
This formula is useful when there are multiple stable fixed points of equations (\ref{MSE_Q}) and (\ref{MSE_M}).
In such a situation the one with the greatest log-likelihood $\phi$ is
the Bayes optimal one.

\subsection{AMP for $U V^T$ decompositions}

In this section we complete the presentation of the AMP algorithm and
the state evolution by stating it for the $U V^\top$ model (\ref{modeluv}).
We remind that $U$ and $V$ are of size $n \times r$ and $m \times r$
respectively and that we consider the following limit.
$n \rightarrow \infty$, $m = \alpha n$, $\alpha = \Omega(1)$, $r = \omega(1)$.
Each row of $U$ and $V$ has been sampled from a probability
distribution $P^{\rm prior}_{u}(u_i)$ and $P^{\rm prior}_{v}(v_i)$
(for simplicity of notation we will omit the upper index "prior" in
this section). Distributions $P_u$ and $P_v$ have their corresponding
input functions $f_u(A,B)$,  $f_v(A,B)$ and their normalizations
${\cal Z}_u(A,B)$, ${\cal Z}_v(A,B)$, defined according to (\ref{DefineFuncF}).
The matrix $W$ is observed through an output channel defined by some
probability $P_{\rm out} (y_{ij}|w_{ij})$. 

The AMP algorithm for estimating $U$ and $V$ from $Y$ is as follows
\begin{eqnarray}
&B^t_{u,i}= \frac{1}{\sqrt{n}}
 \sum\limits_{k=1}^m S_{ik} v^t_{k} - 
 \frac{1}{n\Delta}\left(
 \sum\limits_{k=1}^m \sigma_{v,k}^{t}\right) u^{t-1}_{i}\, ,
\\
&A^t_u = \frac{1}{n \Delta} \sum\limits_{k=1}^m  v^t_{k} {v^t_{k}}^{\top   }\, ,
\\
&u^{t+1}_{i} = f_u(A^t_{u},B^t_{u,i})
\\
&\sigma^{t+1}_{u,i} = \left(\frac{\partial f_u}{ \partial B} \right)(A^t_{u},B^t_{u,i}) \, ,
\\
&B^t_{v,j} = \frac{1}{\sqrt{n}} \sum\limits_{l=1}^n
S_{lj} u_l^t
- 
\frac{1}{n \Delta}\left(
 \sum\limits_{l=1}^n \sigma_{u,l}^t\right) v^{t-1}_{j}\, ,
\\
&A^t_{v} = \frac{1}{n \Delta} \sum\limits_{l=1}^n u^t_{l}{u^t_{l}}^\top   \, ,
\\
&v^{t+1}_{v,j} = f_v(A^t_{v},B^t_{v,j}) \, ,
\\
&\sigma^{t+1}_{v,j} = \left(\frac{\partial f_v}{ \partial B} \right)(A^t_{v},B^t_{v,j}) \, .
\end{eqnarray}
The score matrix $S$ and the effective noise
$\Delta$ were again computed from the output channel
following (\ref{Y_eff}) and (\ref{Noise_Prop_3}).

The Bethe expression for the log-likelihood is obtained from the fixed
point as
\begin{multline}
\phi = \frac{1}{n} \Big\{ \sum\limits_{i=1}^n \log\[[ {\cal
  Z}_u(A_u,B_{u,i})\]]+\sum\limits_{ j=1}^m \log\[[
{\cal Z}_v(A_v,B_{v,j})\]]\Big\}   \\-
\frac{1}{n}\sum\limits_{1 \leq i \leq n,1 \leq j \leq m}
\log(\stackrel{\sim}{Z}_{ij})\, ,
\end{multline}
where ${\cal Z}(A,B_i)$ is again the normalisation from (\ref{DefineFuncF}) and
{\small
\begin{equation}
\log(\stackrel{\sim}{Z}_{ij})={\rm
  Tr}\left[\frac{u_i}{\sqrt{n}}
  S_{ij} v_j^\top - \frac{u_i u_i^\top \sigma_{v,j} + v_j v_j^\top \sigma_{u,i}}{\Delta n}    - \frac{ u_i u_i^\top   }{2 \Delta n}v_j v_j^\top
\right] \, .
\end{equation}
}

To write the state evolution equations we 
introduce the order parameters
\begin{align}
Q_u^{t} = \frac{1}{n}\sum\limits_{1 \leq i \leq n} u_i^t {u_i^t}^\top
\, ,
\quad 
M_u^{t} = \frac{1}{n}\sum\limits_{1 \leq i \leq n} u_i^t {u_0^t}_i^\top
\, ,
\\
Q_v^t = \frac{1}{m}\sum\limits_{1 \leq j \leq m} v_j^t {v_j^t}^\top
\, ,
\quad
M_v^t = \frac{1}{m}\sum\limits_{1 \leq j \leq m} v_j^t {v_0^t}_j^\top
\, ,
\end{align}
for which holds 
{ \small
\bea
&Q_u^{t+1} = \mathbb{E}_{P_{u_0},P_{\xi_v}}
\left[ {f_u{( \frac{\alpha Q_v^t}{\Delta},\frac{\alpha M_v^t}{\Delta}u_0 + \sqrt{\alpha}\xi_v)}
    f_u^{\top   }{(.,.)}} \right]\, ,
\label{MSE_Q_u}
\\
&M_u^{t+1} = \mathbb{E}_{P_{u_0},P_{\xi_v}}
\left[ {f_u{(\frac{\alpha Q_v^t}{\Delta},\frac{\alpha M_v^t}{\Delta}u_0 + \sqrt{\alpha}\xi_v)}
    u_0^\top    }\right]\, ,
\label{MSE_M_u}
\\
&Q_v^{t+1} = \mathbb{E}_{P_{v_0},P_{\xi_u}}
\left[ {f_v{(\frac{ Q_u^t}{\Delta},\frac{ M_u^t}{\Delta}v_0 + \xi_u)}
    f_v^{\top   }{(.,.)}} \right]\, ,
\label{MSE_Q_v}
\\
&M_v^{t+1} = \mathbb{E}_{P_{v_0},P_{\xi_u}}
\left[ {f_v{(\frac{Q_u^t}{\Delta},\frac{M_u^t}{\Delta}v_0 + \xi_u)}
    v_0^\top   } \right]\, ,
\label{MSE_M_v}
\eea  
}
\noindent where $\xi_u$ and $\xi_v$ are Gaussian random variables of mean $0$ and covariance $Q^t_u/\Delta$ and $Q^t_v/\Delta$.
At the fixed point we compute the Bethe free energy as
\begin{multline}
\phi = \alpha\mathbb{E}_{P_{u_0},P_{\xi_v}}  \left[ \log {\cal
         Z}_u(\frac{\alpha Q_v}{\Delta},\frac{\alpha M_v u_0}{\Delta}+\sqrt{\alpha} \xi_v) \right] \\ + \mathbb{E}_{P_{v_0},P_{\xi_u}}  \left[ \log {\cal
         Z}_v(\frac{Q_u}{\Delta},\frac{M_u v_0}{\Delta}+ \xi_u) \right]
\\
- \frac{\alpha{\rm Tr}(M_u M_v^{\top   })}{\Delta} + \frac{\alpha{\rm
  Tr}(Q_u Q_v^{\top   })}{2 \Delta}  \, . \label{Bethe_SE_uv}
\end{multline}

\section{Phase diagrams and examples for clustering}

In this section we illustrate our findings on the example of
clustering into $r$ equally sized groups. In this case the $r$-dimensional rows of
the matrix $X$ take the form of 
$
\left(
0,
\hdots,
1,
\hdots,
0
\right)
$
with a $1$ located at only one of the $r$ coordinates. The matrix $K$
is an identity. The elements of the matrix $W$ are then $w_{ij} = 1/\sqrt{n}$ if
$x_i=x_j$ and $w_{ij}=0$ otherwise. 
Correspondingly the prior $P_{\rm prior}(x)$ is the probability
distribution that picks with probability $1/r$ one of the above $r$
vectors.
The function $f(A,B)$ defined via (\ref{DefineFuncF}) is in this case 
\begin{equation}
f_i(A,B) = \frac{\exp\left(B_i - \frac{A_{ii}}{2} \right)}{
  \sum\limits_{1 \leq k \leq r} \exp\left(B_k - \frac{A_{kk}}{2}
  \right)}\, .
\end{equation}
We will consider two different output channels 
\begin{itemize}
   \item{Gaussian additive noise $P_{\rm out}(y_{ij}|w_{ij}) =
       \exp[-(y_{ij}-w_{ij})^2/(2\Delta)]/\sqrt{2\pi \Delta} $. In this case
       the problem is interpreted as localization of submatrices
       having mean $1/\sqrt{n}$ in an otherwise zero-mean matrix.}
   \item{Stochastic block model channel where $Y$ is its (dense) adjacency
       matrix created as $P_{\rm out}(y_{ij}=1|w_{ij}) = p_{\rm out}
       + \mu w_{ij}$ and $P_{\rm out}(y_{ij}=0|w_{ij}) = 1-p_{\rm out} - \mu
       w_{ij}$.  That is, nodes from different groups are connected with
       probability $p_{\rm out}$, and nodes from the same group with
       probability $p_{\rm in} = p_{\rm out} + \mu/\sqrt{n}$. Both $p_{\rm
         out}=\Omega(1)$, and $\mu = \Omega(1)$. 
        The Fisher score matrix has elements $S_{ij}= \mu/p_{\rm out}$ where
        $Y_{ij}=1$, and $S_{ij}=-\mu/(1-p_{\rm out})$ where $Y_{ij}=0$. The inverse Fisher information of the
stochastic block model channel is given as 
\begin{equation}
\Delta = \frac{p_{\rm out}(1-p_{\rm out})}{\mu^2}\, . 
\end{equation}
In the large $n$ limit the resulting graph will have average degree $n
p_{\rm out}$. 
}
\end{itemize}

Bayes-optimal inference in the stochastic block model was studied in
the case of sparse networks (bounded average degree) \cite{decelle2011asymptotic}. For the case
of dense networks (linear average degree) and $\mu = \Omega(1)$ the Bayes-optimal MMSE and
corresponding phase transitions are an original contribution of the
present paper.

\subsection{MMSE and phase transitions for the symmetric clustering}


The state evolution from section \ref{SE} simplifies for the
case of symmetric clustering in the setting above. One notices that if $Q^t$ and
$M^t$ are of the symmetric form
\begin{equation}
Q^t = M^t = \frac{a^t}{r^2} J_r + \frac{b^t}{r} I_r
\end{equation}
then $Q^{t+1}$ and $M^{t+1}$ will be of the same form.
Here $J_r$ is a $r\times r$ matrix filled with ones, and $I_r$ is a
$r\times r$ identity matrix. 
Moreover since the sum of elements of the vectors $x$ is always $1$
we have the additional property that 
\begin{eqnarray}
a^t + b^t = 1\, , \quad \quad 
0 \leq a^t,b^t \leq 1 \, . 
\end{eqnarray}
Therefore in the end we only have one parameter $b$ in the state evolution,
$b^t = 1$ means that we have perfectly reconstructed the communities,
$b^t = 0$ is when there is no information and the estimator for every variables is $\left(\frac{1}{r},\cdots,\frac{1}{r}\right)$.

The state evolution equation (\ref{MSE_Q}) can then be used to derive
the evolution of the parameter $b$ under iterations. It follows that 
\begin{equation}
b^{t+1} = {\cal M}_r\left(\frac{b^t}{\Delta}\right) \, ,\label{b_Equation_Simplified}
\end{equation}
where the function ${\cal M}$ is 
{\small
\begin{equation}
{\cal M}(x) = \frac{r}{r - 1}\left[\int
\frac{\exp \left( \frac{x}{r} + u_1 \right)}
{
\exp \left( \frac{x}{r} + u_1 \right) + \sum\limits_{i = 2}^r \exp \left( u_i \right)
}
\prod\limits_{i=1}^r  {\cal D} u_i  -\frac{1}{r}  \right] \, ,\label{SE_b_Community}
\end{equation}
}
where we introduced a Gaussian measure 
\begin{equation}
   {\cal D} u_i =  {\rm d} u_i  \frac{\sqrt{r}}{\sqrt{2 \pi x}}
   \exp{\left(\frac{-r u_i^2}{2 x} \right)}  \, .
\end{equation}

We now observe that the above state evolution equation has always the uniform or
uninformative fixed point $b_u = 0$. It is crucial to study the
stability of this fixed point under iterations of (\ref{b_Equation_Simplified}).
Depending on the number of groups $r$ and the noise parameter $\Delta$ there may be another stable fixed point of (\ref{b_Equation_Simplified}).
When it exists we will call it $b_{\rm far}$ (it obviously depends on $\Delta$ and $r$).
We expand (\ref{SE_b_Community}) around $b_{u}$ to get
\begin{equation}
b^{t+1} = \frac{b^t}{\Delta r^2} + \frac{r-4}{2 \Delta^2 r^4} {b^t}^2
+ O({b^t}^3)\, . \label{SE_b_Community_Expansion}
\end{equation}
From this equation we see that for
\begin{equation}
\Delta < \Delta_{c}=\frac{1}{r^2}\,   \label{deltac}
\end{equation}
the uniform fixed point will be unstable and the iteration will
converge away from it. 

If we translate condition (\ref{deltac}) back to the parameters of the
stochastic block model we obtain that inference with AMP is possible
for 
\begin{equation}
           |p_{\rm in} - p_{\rm out}| > \frac{1}{\sqrt{n}} r
           \sqrt{p_{\rm out}(1-p_{\rm out})} \, .
\end{equation}
This is the same condition as known from the sparse SBM
\cite{decelle2011asymptotic} and from standard spectral methods
\cite{baik2005phase,nadakuditi2012graph}. Our contribution to these results is the
Bayes-optimal value of the detection error, which is considerably
better than the error obtained from the spectral algorithms evaluated
in \cite{nadakuditi2012graph}. Note e.g. that the error in the
spectral algorithm is always continuous at the transition, whereas
the Bayes-optimal error presents a discontinuity at the transitions
for $r>4$ (see below). 

By looking at the second order term in (\ref{SE_b_Community_Expansion}) we can get
some information about where the iterations will converge. 
\begin{itemize}
\item If $r< 4$, $\frac{r-4}{2 \Delta^2 r^4} {b^t}^2$ is
  negative and close to $\Delta_c$ we will converge to 
$
b_{\rm far}(\Delta) = 2(\Delta_c - \Delta)/(4-r) + O[(\Delta_c -
\Delta)^2]
$.
In this case the fixed point is a continuous function of the noise.  
\item If $r > 4$, things are different. Because the second
  derivative of (\ref{SE_b_Community_Expansion}) is positive then it
  is impossible for $b_{\rm far}$ to go to zero continuously as
  $\Delta$ goes to $\Delta_c$ (a simple plot of
  (\ref{SE_b_Community_Expansion}) should convince the reader of this
  fact). There is therefore a jump in the value of the fixed point as $\Delta$ crosses $\Delta_c$. This is the signature of a first order phase transition.
\end{itemize}

In the case of first order phase transition there are always three
transitions to discuss  
\begin{equation}
\Delta_{\rm c} = \frac{1}{r^2} < \Delta_{\rm static}<\Delta_{\rm
  spinodal}\, .
\end{equation}
The different phases have the following properties: 
\begin{itemize}
\item Undetectable phase without clusters, $\Delta_{\rm spinodal} < \Delta$.  Eq. (\ref{b_Equation_Simplified}) has only the uniform fixed
  point. Bayes optimal inference does not provide any information
  about the labeling of the nodes.  
\item Undetectable phase with clusters, $\Delta_{\rm static}<\Delta<\Delta_{\rm spinodal}$. Eq. (\ref{b_Equation_Simplified}) has two
stable fixed point $b_{\rm u}$ and $b_{\rm far}$. But $b_{\rm far}$
has a lower log-likelihood than $b_{\rm u}$ therefore the
Bayes-optimal estimator is still not correlated with the planted solution.
\item Detectable but hard, $\Delta_{c}<\Delta<\Delta_{\rm static}$. The fixed point $b_{\rm far}$ now has a larger
log-likelihood than $b_{\rm u}$. Bayes optimal estimator will find a
configuration well correlated with the planted one, but the AMP
algorithms will not. Analogous phase appears in many other inference
problems and we conjecture that all known polynomial
algorithms will fail in this phase. 
\item Easy phase, $\Delta<\Delta_{c}$. The uniform fixed point $b_{\rm
    u}$ is unstable and the AMP algorithm will converge to $b_{\rm
    far}$. The problem is said easy since we have a polynomial
  algorithm that (at least is conjectured to) asymptotically reaches optimal reconstruction.
\end{itemize}

A theoretical analysis of $\Delta_{\rm spinodal}(r)$ and $\Delta_{\rm
  static}(r)$ is also possible as $r\to\infty$. It relies on analysis of
which exponential dominates the others in (\ref{SE_b_Community}), for
details see appendix D.
We find
\begin{eqnarray}
\Delta_{\rm spinodal} = \frac{1}{2 r \ln(r)} [1+ o(1)] \, ,\label{ExpansionSpinodal}\\
\Delta_{\rm static} = \frac{1}{4 r \ln(r)} [1+ o(1)]\, .\label{ExpansionStatic}
\end{eqnarray}
The most notable remark about these results is that in the limit of
large rank the gap between what is statistically possible $\Delta_{\rm
static}$ and what is algorithmically tractable $\Delta_c=1/r^2$ is
different even in its order. Note also that the present phase
transitions are equal to critical temperatures in the Potts glass
model as computed in \cite{gross1985mean,caltagirone2012dynamical}, this is because of
the intimate relation between random and planted models~\cite{KrzakalaZdeborova09}. Note that these works also derived rank
$r=4$ as the point where the second order phase transitions changes
into a first order one.

\subsection{Phase diagrams}
In this section we illustrate the behavior of the fixed points of the AMP algorithm and of the state evolution.
In all the presented experiments we iterate the corresponding
equations till convergence. To investigate the MMSE and the phase
transitions we can initialize the AMP algorithm and the SE equations in two different ways:
\begin{itemize}
\item Uniformative initialization: for the SE equations this means
  $b^{t=0} = \delta$, where $\delta$ is very small. For the AMP
  algorithm this means $a_i=(\frac{1}{r},\hdots,\frac{1}{r})^\top +
  \delta_i$ and $v_i^t = I_r/r$. This is the initialization with which
  we would work if we were working with real data.
\item Informative initialization: for the SE equations this means
  $b^{t=0} = 1$. For the AMP algorithm this means $a_i=x_{i}$ and
  $v_i^t = 0$, i.e. initiate the estimators to be equal to the planted solution that we are trying to recover. 
\end{itemize}

In community detection we usually  evaluate an estimator that
maximizes the number of correctly assigned nodes. For this we need to
take the index of the maximum component of $a_{i}$. For numerical
comparisons between the state evolution and the AMP algorithm we will
rather evaluate the mean-squared error ${\rm MSE}=1-Q$, where $Q$ is the order parameter~(\ref{op_Q}). 

\begin{figure}[h]
\begin{center}
\includegraphics[scale=1]{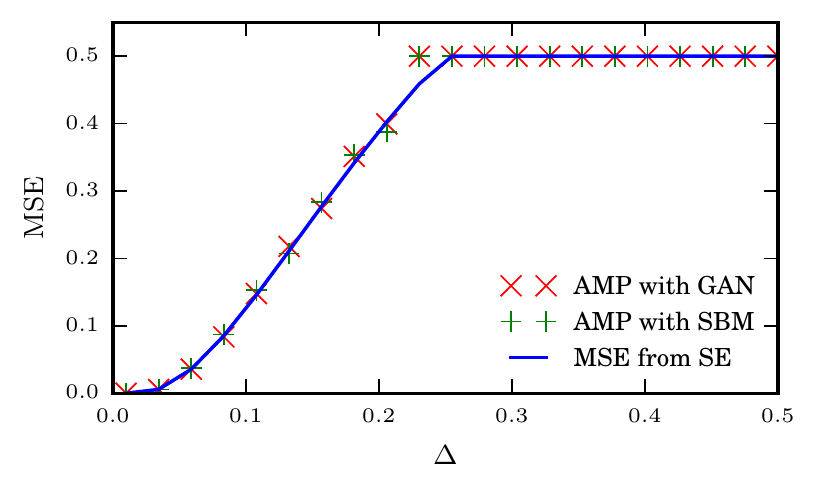}
\end{center}
\caption{The MSE of the reconstruction of  communities in the rank
  $r=2$ case. We present the result of AMP for two different channels
  the Gaussian additive noise (GAN, red crosses) and the stochastic block model
  (SBM, with $p_{\rm out} = 0.5$, green crosses) compared to the
  theoretical value from the state
  evolution (SE, blue line). The AMP simulations were run on a single instance of size $n=20000$.}
\label{Fig_MSE_r=2}
\end{figure}

In figure \ref{Fig_MSE_r=2} we illustrate the fact that in the
large $n$ limit different output channels having the same inverse
Fisher information $\Delta$ have equivalent performance. We plot the
MSE obtained from the state evolution, compare to the one obtained
from the AMP algorithm on a single instance of size $n=20000$ with
either a Gaussian additive noise channel or the stochastic block model
channel. Indeed the three cases agree. 

\begin{figure}[h]
\begin{center}
\includegraphics[scale=1]{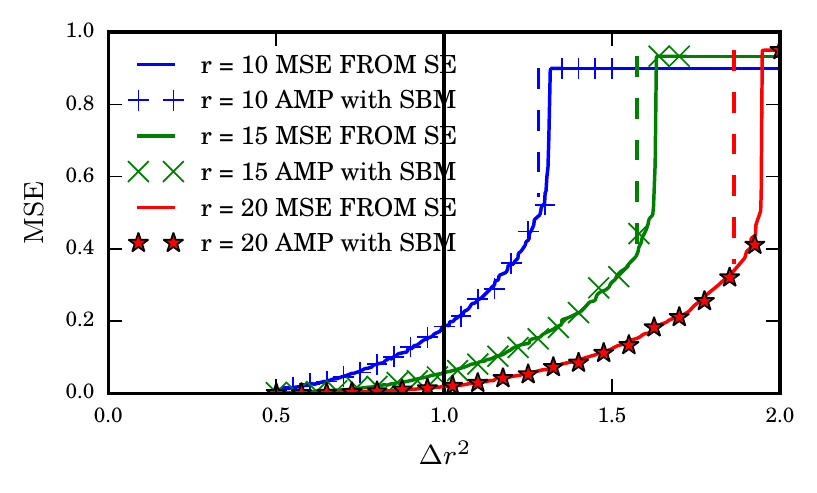}
\end{center}
\caption{
We plot MSE deduced from state evolution (lines) and from AMP (marks) for different values of rank
$r$ as a function of $\Delta r^2$ (this rescaling is for visual
reasons). The vertical full black line is $\Delta_c$ for all the cases. The
vertical dashed colored lines are $\Delta_{\rm static}$ and the full lines
correspond to the MSE obtained from the informative initializations
and have discontinuities at $\Delta_{\rm spinodal}$.}
\label{Fig_FirstOrderTransition}
\end{figure}

In figure \ref{Fig_FirstOrderTransition} we depict the first order
phase transition that occurs for $r>4$. We present the MSE obtained
from density evolution as function of $r^2 \Delta$ for various ranks
$r=10,15,20$. From the uninformative initialization the MSE jumps to $1-1/r$ at
$\Delta_c = 1/r^2$. From the informative initialization the jump occurs
at $\Delta_{\rm spinodal}>\Delta_c$ and by comparing the Bethe free energy
of the two fixed points we evaluate the $\Delta_{\rm static}$. We
observe that already for these values of the rank the gap between the
information theoretical and computationally tractable performance,
i.e. between $\Delta_c$ and $\Delta_{\rm static}$ is very large. 
Another interesting observation we made concerns the value of the MSE at the
$\Delta_c=1/r^2$ transition. The reconstruction is very close to exact even at the phase
transition itself. More precisely it decreases roughly exponentially with the rank
$r$, being close to $10^{-10}$ for rank $r=100$.  

\begin{figure}[h]
\begin{center}
\includegraphics[scale=1]{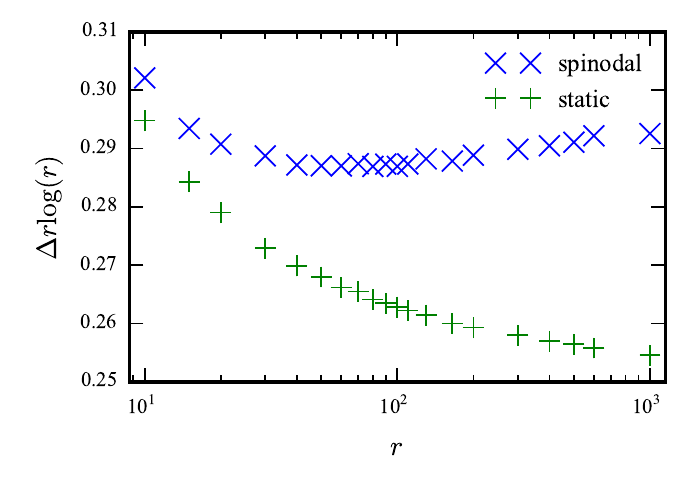}
\end{center}
\caption{
We plot  $\Delta r \log{r}$ for the static and spinodal phase
transitions obtained from the state evolution using the protocol
described in Appendix~C. We rescale the $\Delta$ in
this way to compare with the large rank expansion in (\ref{ExpansionSpinodal}) and (\ref{ExpansionStatic}).} 
\label{Fig_PhaseDiagramm}
\end{figure}

In figure \ref{Fig_PhaseDiagramm} we plot the values of $\Delta_{\rm
  spinodal}$ and $\Delta_{\rm static}$ rescaled by $r\log{r}$ as a function of the rank
$r$. As the rank grows $r\to \infty$ the static line will converge to
$1/4$ and the spinodal one to $1/2$ (\ref{ExpansionSpinodal}-\ref{ExpansionStatic}). Whereas the static transition
converges nicely to its asymptotic value, for the spinodal transition we
are still quite far from the asymptotic regime.





\section{Conclusions and perspectives}

We considered here the problem of estimation of a low rank matrix that
was observed element-wise trough an arbitrary noisy channel. Using
approximate message passing and its state evolution we compute the
Bayes-optimal MMSE in this problem. The most interesting and,
comparing to previously studied cases, surprising conclusion is that
the MMSE depends on the output channel only trough its Fisher
information. 

As an example of the considered setting we study the problem of
clustering in dense stochastic block model with the difference between
probabilities of connection within and between communities scaling as
$O(1/\sqrt{n})$. We evaluate the phase transitions and MMSE in this
regime, unveiling a wide region of computational hardness. 

Both the AMP algorithms and the state evolution apply to a number of
other setting considered in the literature, which is a natural
direction of future work.

\section{Acknowledgements}
We thank Madhu Advani for clarifying discussion about Fisher
information. The research leading to these results has received funding from the
European Research Council under the European Union's $7^{th}$
Framework Programme (FP/2007-2013)/ERC Grant Agreement 307087-SPARCS.

\bibliographystyle{IEEEtran}
\bibliography{IEEEabrv,refs}


\section*{Appendix A}

To derive the state evolution let us define 
\begin{equation}
G_{il} = \frac{K}{\sqrt{n}}
S_{il} {a_{l \rightarrow i
    l}^{t}}.
\end{equation}
We then have $B_i^t= \sum\limits_{l \neq i} G_{il}$. We use the central limit theorem to find the distribution of $B_i$ in order to do that one needs to compute, $\mathbb{E}_{Y}(G_{il})$ and ${\rm Cov}(G_{il}G_{ik}^\top)$.
Let us show how to compute the mean
\begin{equation}
\mathbb{E}(G_{il}) = \int {\rm d} y P_{\rm out}\left(y|w_{il}\right)\frac{K}{\sqrt{n}}
\frac{\partial g}{\partial
    w}\Big|_{y,0} {a_{l \rightarrow i
    l}^{t}} \label{MeanG_ij}.
\end{equation}
We then expand the probability $P_{\rm out}(y|w)$ around 0
\begin{equation}
P_{\rm out}\left(y|w\right) = P_{\rm out}(y|0) \left(1 + w \frac{\partial g}{\partial
    w}\Big|_{y,0}\right) \, .\label{Perturbation_P(Y)}
\end{equation}
Using (\ref{Perturbation_P(Y)}) in (\ref{MeanG_ij}) and using ({\ref{Noise_Prop_1}}) and ({\ref{Noise_Prop_1}})
we get
\begin{equation}
\mathbb{E}(G_{il}) = \frac{K}{n \Delta} a_j x_j^\top K  x_l
+ O\left(n^{-\frac{3}{2}}\right)\, ,
\end{equation}
where $a_i$ is the mean of $a_{i \rightarrow il}$ with respect to $y_{ij}$.
Using (\ref{Perturbation_P(Y)}) and the fact that $a_{l \rightarrow
  il}$ and $a_{k \rightarrow ik}$ are independent random variables we
can compute the second moments of the $G_{ij}$ as
\begin{equation}
{\rm Cov}(G_{il},G_{ik}) = \delta_l^k \frac{K a_i a_i^\top K}{n \Delta} ,
\end{equation}
From the central limit theorem we deduce.
\begin{equation}
B_i^t = \frac{K M^t K {x}_i}{\Delta} + W_i \, , \label{Distribution_B_i}
\end{equation}
where $W_i$ is a Gaussian random variable of mean 0 and of covariance
$K Q^t K/\Delta$.

\section*{Appendix B}

The Bethe free energy in the case where all the factors are between two
variables reads as
\begin{equation}
\Phi =n\phi= \sum\limits_{1 \leq i \leq n} \mathbb{F}_i -
\sum\limits_{1 \leq i\leq j\leq n} \mathbb{F}_{ij} \, ,
\end{equation}
where
\begin{equation}
\mathbb{F}_i = \log\left[\int {\rm d}x_i P_{\rm prior} \prod_{j \neq i} m_{ij\rightarrow i}(x_i)\right]
\end{equation}
and
\begin{equation}
\mathbb{F}_{ij} = \log\left[\int {\rm d}x_i {\rm d}x_j P_{\rm out}\left(y_{ij}|w_{ij}\right) n_{i \rightarrow ij}(x_i)n_{j \rightarrow ij}(x_j)\right].
\end{equation}
Using (\ref{ExpansionBP}) we find
\begin{equation}
\sum\limits_{1 \leq i \leq n} \mathbb{F}_i= \sum\limits_i \log {\cal
  Z}(A,B_i) \, .\label{VariableEnergy}
\end{equation}
For the second term we expand once again $P_{\rm out}\left(y_{ij}|w_{ij}\right)$ around 0 and then take the mean with respect to each message.
We then take the log and find
{
\begin{eqnarray}
\nonumber &\mathbb{F}_{ij} = g(y_{ij},0) + \frac{a_{i \rightarrow ij}^\top K a_{j \rightarrow ij}}{\sqrt{n}}S_{ij}-&
\\
\nonumber
&\frac{S_{ij}^2}{2n}{\rm Tr}\left[K a_{i \rightarrow i j}^t {a_{i \rightarrow i j}^{t}}^\top   
K
a_{j \rightarrow i j}^t {a_{j \rightarrow i j}^{t}}^\top   
\right]+
\\
\nonumber
&\frac{h(y_{ij})}{2n}     {\rm Tr }\left[K\left(v_{l \rightarrow i l}^t +
a_{i \rightarrow i j}^t {a_{i \rightarrow i j}^{t}}^\top   
\right) \right. \\ & \times \left.  K \left(v_{l \rightarrow i l}^t +
a_{j \rightarrow i j}^t {a_{j \rightarrow i j}^{t}}^\top   \label{F_ij_Complicated} 
\right)\right] \, .
\end{eqnarray}}
The TAPification procedure that is not detailed here gives us
\begin{equation}
a_{i\rightarrow ij} = a_i - \frac{S_{ij}a_j}{\sqrt{n}} + O\left(\frac{1}{n} \right) \label{TAP_Begin} \, .
\end{equation}
We also remove the term $g(y_{ij},0)$ since it is a constant that does
not depend on the messages.
We replace (\ref{TAP_Begin}), (\ref{Noise_Prop_2}) and (\ref{Noise_Prop_3}) in (\ref{F_ij_Complicated}) and we get
\begin{multline}
\mathbb{F}_{ij} = \frac{a_{i}^\top K a_{j}}{\sqrt{n}}S_{ij}
-\frac{1}{2n\Delta}{\rm Tr}\left[K a_{i}^t {a_{i}^{t}}^\top   
K a_{j}^t {a_{j}^{t}}^\top
\right]\\
\frac{1}{\Delta n}{\rm Tr}\left[K a_i a_i^\top K v_j + K a_j a_j^\top K v_i  \right]\, .\label{F_ij_Simplified}
\end{multline}
Since the sum is on all undirected links $ij$ we do the sum on all directed link and divide by $2$
\begin{multline}
\Phi = \sum\limits_{1 \leq i \leq n} \log {\cal Z}(A,B_i) -\\ \frac{1}{2}\sum\limits_{1 \leq ij\leq n} \frac{a_{i}^\top K a_{j}}{\sqrt{n}}S_{ij}
-\frac{1}{2n\Delta}{\rm Tr}\left[K a_{i}^t {a_{i}^{t}}^\top   
K a_{j}^t {a_{j}^{t}}^\top
\right]-
\\
{\rm Tr}\left[\frac{K}{n \Delta}\left(\sum\limits_{1 \leq i\leq n} a_i a_i^\top\right)K\left(\sum\limits_{1 \leq i\leq n} v_i \right) \right]
\end{multline}

To get the Bethe-free-energy in the large $n$ limit we compute the mean of
$\frac{a_{i \rightarrow ij}^\top K a_{j \rightarrow ij}}{\sqrt{n}}$ with respect to the noise.
The main idea remains the same we compute the mean with respect to a perturbation of $P_{\rm out}$
\begin{multline}
\mathbb{E}\left(\frac{S_{ij}a_{i \rightarrow ij}^\top K a_{j
      \rightarrow ij}}{\sqrt{n}} \right) = \\ \int {\rm d}y P_{\rm
  out}(y|0) \left(1 + S_{ij}w_{ij}\right)\frac{S_{ij}a_{i \rightarrow ij}^\top K a_{j \rightarrow ij}}{\sqrt{n}} 
\end{multline}
\begin{equation}
\mathbb{E}\left(\frac{S_{ij}a_{i \rightarrow ij}^\top K a_{j
      \rightarrow ij}}{\sqrt{n}} \right) =  \frac{K x_i a_i^\top K a_j
  x_j^\top}{n\Delta } \, .\label{MeanFij}
\end{equation}
To compute the mean of (\ref{VariableEnergy}) we use the fact that we know the distribution of the $B_i$ (\ref{Distribution_B_i}).
We replace (\ref{MeanFij}) in (\ref{F_ij_Complicated}) and we get
\bea
     \phi &= \mathbb{E}_{P_x,P_W}  \left[ \log {\cal
         Z}(\frac{KQK}{\Delta},\frac{KM Kx}{\Delta}+W) \right]
\nonumber \\
&- \frac{1}{2 \Delta} {\rm Tr}(KMK M^{\top   }) + \frac{1}{4 \Delta} {\rm
  Tr}(KQK Q^{\top   }) \, . \label{Bethe_SE_II}
\eea

\section*{Appendix C}

To locate the phase transitions $\Delta_{\rm static}$ and $\Delta_{\rm spinodal}$ we make a couple of remarks about
the state evolution. 
First we remark that for $\forall x \in \mathbb{R}^+$
$b = {\cal M}_r(x)$ is a fixed point of (\ref{b_Equation_Simplified}) for
$\Delta = {{\cal M}_r(x)}/{x}$. By definition $\Delta_{\rm spinodal}$ is the greatest $\Delta$ for which $b_{\rm far}$ exist. Therefore
\begin{equation}
\Delta_{\rm spinodal} = \max_{x  \in
  \mathbb{R}^+}\left\{\frac{{\cal M}_r(x)}{x} \right\}\, .\label{FormulaSpinodal}
\end{equation}
To find the $\Delta_{\rm static}$ one can notice that by taking the derivative
with respect to $Q$ and $M$ of (\ref{Bethe_SE}) one finds a combination of (\ref{MSE_Q}) and (\ref{MSE_M}).
Therefore we have
\begin{equation}
\phi(b_2,\Delta) - \phi(b_1,\Delta) =  \frac{r-1}{2 r^2
  \Delta}\int\limits_{b_1}^{b_2} {\rm d}u {\cal
  M}_r\left(\frac{u}{\Delta}\right)-u \, .\label{ProtocolTransition}
\end{equation}
We deduce a way to compute $\Delta_{\rm static}$ as
{\small
\begin{equation}
\Delta_{\rm static} = \max_{x  \in
  \mathbb{R}^+}\left\{\frac{{\cal M}_r(x)}{x}, {\rm s.t.}\,
  \int\limits_{0}^{x} {\rm d}u {\cal M}_r(u)=\frac{x{\cal M}_r(x)}{2}
\right\}\, . \label{FormulaStatic}
\end{equation}}
To compute $\Delta_{\rm static}$ we evaluate ${\cal M}_r$ on a whole interval then for each $x$ draw a
line between point $(0,0)$ and $(x,H(x))$. We then compute the area
between ${\cal M}_r$ and this line. When this aera is zero then ${\cal M}_r(x)/{x}$ gives us $\Delta_{\rm static}$.

\section*{Appendix D}

In this appendix we explain the derivation of (\ref{ExpansionSpinodal}), (\ref{ExpansionStatic}).
To do this let us study the function ${\cal M}_r(x)$ where we take $x =
\beta r \log(r)$, where $\beta = \Omega(1)$.
The important part of ${\cal M}_r$ is this integral
\begin{equation}
\int
\frac{\exp \left( \frac{x}{r} + u_1 \right)}
{
\exp \left( \frac{x}{r} + u_1 \right) + \sum\limits_{i = 2}^r \exp \left( u_i \right)
}
\prod\limits_{i=1}^r  {\cal D} u_i\, .
\end{equation}
The important variables to look at are
\begin{eqnarray}
F_1 = \exp \left( \frac{x}{r} + u_1 \right)\, , \label{M_r1}
\\
F_2 = \sum\limits_{i = 2}^r \exp \left( u_i \right). \label{M_r2}
\end{eqnarray}
If $F_1$ dominates $F_2$ as $r \rightarrow +\infty$ then ${\cal M}_r = 1$, otherwise if
$F_2$ dominates $F_1$ then ${\cal M}_r = 0$.


To estimate $F_1$, $F_2$ let us notice that with high probability the maximum value will be of order
$\sqrt{2 \beta \log(r)}$. This is a general property of Gaussian variables that the maximum of~$r$ independent Gaussian variables of variance $\sigma^2$ is of order $\sigma \sqrt{2 \log(r)}$.
We can therefore compute the mean of $F_2$ all while conditioning on
the fact that all of the $u_i$ are smaller than $\sqrt{2 \beta
  \log(r)}$. This allows us to compute the typical value of $F_1$ as $
F_1 \sim r^\beta $. For $F_2$ we obtain:  
when $\beta < 1/2$ then
$F_2 \sim r^{\beta + \frac{1}{2}}$, and when 
 if $\beta > 1/2$ then $F_2 \sim r^{\sqrt{2 \beta}}$.
We have to look at which of the $F_1$ or $F_2$ has a higher exponent.
In these computation we can therefore just assume that
\begin{equation}
{\cal M}_r(\beta r \log(r)) = \mathbb{1}\left(\beta > 2 \right)\, .
\end{equation}
Now let us remind (\ref{FormulaSpinodal}) while keeping $x = \beta r \log(r)$
\begin{equation}
\Delta_{\rm spinodal} = \max \left\{\frac{\mathbb{1}\left(\beta > 2 \right)}{ \beta r \log(r)}, \beta \in \mathbb{R}^+\right\} = \frac{1}{2 r \log(r)}
\end{equation}

To get the static transition we use (\ref{FormulaStatic}).
Let us find the $\beta$ that satisfies equation (\ref{FormulaStatic}) we get
\begin{equation}
\beta r \log(r) \left[1 - \frac{2}{\beta}\right ] = \frac{\beta r
  \log(r)}{2}\, .
\end{equation}
We deduce $\beta = 4$ and therefore 
\begin{equation}
\Delta_{\rm static} = \frac{1}{4 r \log(r)}\, .
\end{equation}

\end{document}